\documentclass[useAMS,usenatbib,usegraphicx]{mn2e}
\usepackage{natbib}
\usepackage{color}
\usepackage{mathrsfs}
\usepackage{times,xspace,amssymb}
\usepackage{url}
\usepackage{hyperref}
\usepackage{amsmath}
\usepackage{ulem}

\bibpunct{(}{)}{;}{a}{}{,}

\def\mnras{MNRAS}
\def\apj{ApJ}

\def\aanda{A\&A}

\def\prb{Phys. Rev. B}
\def\prl{Phys. Rev. Letters}
\def\araa{ARA\&A}

\def\newastrev{New. Ast. Reviews}
\def\arxiv#1{\href{http://arxiv.org/abs/#1}{arXiv:#1}} 
\def\adspr{Adv. Space Research}
\def\vbh{V_{\bullet}}
\def\v2{V_{200}}

\def\lbh{L_{\bullet}}
\def\rbondi{R_{\rm Bondi}}
\def\mbh{M_{\bullet}}
\def\m5{M_{5}}
\def\cs{c_{\rm s}}
\def\epnt{\epsilon_{\rm nt}}
\def\ent5{\epsilon_{\rm nt, 5}}
\def\msun{M_{\odot}}
\def\nism{n_{_{\rm ISM}}}
\def\mp{m_{\rm p}}
\def\n0{n_{0}}
\def\rsun{R_{\odot}}

\def\vghz{\nu_{_{\rm GHz}}}
\def\gmin{\gamma_{\rm min}}

\def\gmax{\gamma_{\rm max}}
\def\tacc{t_{\rm acc}}
\def\tdyn{t_{\rm dyn}}
\def\xacc{\xi_{\rm acc}}
\def\tc{t_{\rm cool}}

\def\t4{T_{4}}
\def\d1{d_{1}}

\def\b-5{B_{-5}}

\def\g4{\gamma_{4}}
\def\gb{\gamma_{\rm b}}

\def\jv{j_{\nu}}
\def\av{\alpha_{\nu}}
\def\tv{\tau_{\nu}}
\def\lnt{L_{\rm nt}}

\def\mel{m_{\rm e}}
\def\kms{\rm km\,s^{-1}}
\def\kpc{\rm kpc}
\def\mach{\mathcal{M}}
\def\medd{\dot{M}_{\rm Edd}}
\def\mbondi{\dot{M}_{\rm Bondi}}
\def\rout{R_{\rm out}}
\def\vout{V_{\rm out}}
\def\vouta{V_{\rm out,4}}

\def\mout{\dot{M}_{\rm out}}
\def\zetaa{\zeta_{-1}}

\title[Detecting floating black holes]{%
Detecting floating black holes as they traverse the gas disk of the Milky Way}
\author[X. Wang and A. Loeb]{%
Xiawei Wang\thanks{E-mail: xiawei.wang@cfa.harvard.edu}
and Abraham Loeb
\\
Department of Astronomy, Harvard University, 60 Garden St., Cambridge, MA 02138, USA}

\begin{document}
\date{Accepted ... Received...; in original form..}

\pagerange{\pageref{firstpage}--\pageref{lastpage}} \pubyear{2014}

\maketitle

\label{firstpage}
\begin{abstract}
A population of intermediate-mass black holes (BHs) is predicted 
to be freely floating in the Milky Way (MW) halo,
due to gravitational wave recoil, ejection from triple BH systems, or tidal stripping
in the dwarf galaxies that merged to make the MW.
As these BHs traverse the gaseous MW disk, a bow shock forms, 
producing detectable radio and mm/sub-mm
synchrotron emission from accelerated electrons.
We calculate the synchrotron flux to be $\sim \rm 0.01-10\, mJy$ at GHz frequency,
detectable by {\it Jansky Very Large Array}, 
and $\sim 10-100\,\mu\rm Jy$ at $\sim10^{10}-10^{12} \,\rm Hz$ frequencies,
detectable by {\it Atacama Large Millimeter/sub-millimter Array}.
The discovery of the floating BH population will provide insights on 
the formation and merger history of the MW as well as on the
evolution of massive BHs in the early Universe.
\end{abstract}
\section{Introduction}
Galaxies grow through accretion and hierarchical mergers.
During the final phase of the merger of two central black holes,
anisotropic emission of gravitational waves (GW) kicks the BH remnant 
with a velocity up to a few hundreds $\kms$ 
\citep{baker2006, campanelli2007, blecha2008}.
Additionally, BHs can be ejected from triple systems 
\citep{kulkarni2012, hoffman2007},
or result from tidally-stripped cores of dwarf galaxies \citep{bellovary2010}.
For GW recoils, 
the typical kick velocity is large enough for the BHs to escape the shallow gravitational potential of
low-mass galaxies, but smaller than the escape velocity of the MW halo.
This is also the case for triple systems as long as the kick velocity is $<500 \,\kms$.
Consequently, a population of floating BHs formed from
mergers of low-mass galaxies are trapped in the region that eventually makes the 
present-day MW \citep{madau2004, volonteri2005, libeskind2006}.
Previous studies suggested that more than $\sim$ 100 floating BHs 
should be in the halo today,
based on a large statistical sample of possible merger tree histories 
for the MW halo today \citep{oleary2009, oleary2012}.
This population of recoiled BHs is supplemented by BHs
from tidally disrupted satellites of the MW \citep{bellovary2010}.
The discovery of this BH population will provide constraints on the
formation and merger history of the MW as well as the dynamical evolution
of massive BHs in the early Universe.
 
It has been proposed that a compact cluster of old stars from the original host galaxies 
is carried by each floating BH \citep{oleary2009}.
In this Letter, we propose an additional observational signature of floating BHs,
using the MW gas disk as a detector.
As the BHs pass through the MW disk supersonically
they generate a bow shock, which results in
synchrotron radiation detectable at radio and mm/sub-mm frequencies.

The paper is organized as follows.
In \S~\ref{sec:interact}, we discuss the interaction between BHs and the gas in the MW disk.
In \S~\ref{sec:obs}, we calculate the synchrotron radiation from the bow shocks 
produced as the BHs cross the MW disk, 
and discuss the detectability of this radiation.
Finally, in \S~\ref{sec:sum}, we summarize our results and discuss their implications.
\section{Interaction between a floating black hole and the MW disk gas}
\label{sec:interact}

We consider a BH moving at a speed $\vbh$ relative to the interstellar medium (ISM)
of number density $\nism$.
The effective radius of influence of the moving black hole is given by
the Bondi accretion radius:
\begin{equation}
\rbondi=\frac{G\mbh}{\cs^{2}+\vbh^{2}} \approx \frac{G\mbh}{\vbh^{2}} 
=0.01\,M_{5}\,V_{200}^{-2}\; \rm pc\; ,
\end{equation}
where $G$ is Newton's constant, 
$M_{5}=\left(\mbh/10^{5}\msun\right)$ and $V_{200}=\left(\vbh/200\,\kms\right)$.
The sound speed $\cs$ of hydrogen in the ISM is given by 
$\cs = \left(\Gamma P/\rho\right)^{1/2}=11.7\,\t4^{1/2}\, \kms$,
where $\Gamma=5/3$ is the adiabatic index and $\t4=\left(T/10^{4}\,\rm K\right)$.
In the case of a supersonic shock with velocity $V_{\rm sh} \gg \cs$, 
the total mass enclosed within the Bondi radius is given by
$\Delta M_{\rm ISM}=1.3\times 10^{-7}\,\m5^{3}\,\v2^{-6}\,\n0\;\msun$,
where $\n0=\left(\nism/10^{0}\,\rm cm^{-3}\right)$.
The rate of fresh mass being shocked in the ISM is
$\Delta\dot{M}_{\rm ISM}=
3.7\times 10^{-9}\,\m5^{2}\,\v2^{-3}\,\n0\; \msun\,\rm yr^{-1}$.
The total kinetic power can be expressed as,
\begin{equation}
\begin{split}
L_{\rm kin}&=\frac{1}{2}\left(2\pi \rbondi^{2}\nism m_{\rm p}\vbh \right)\vbh^{2}\\
&=4.7\times10^{31}\m5^{2}\,\v2^{-1}\,\n0\;\rm erg\,s^{-1}\; ,
\end{split}
\end{equation}
where $m_{\rm p}$ is the proton mass.
\section{Observational appearance}
\label{sec:obs}
\begin{figure}
\includegraphics[angle=0,width=\columnwidth]{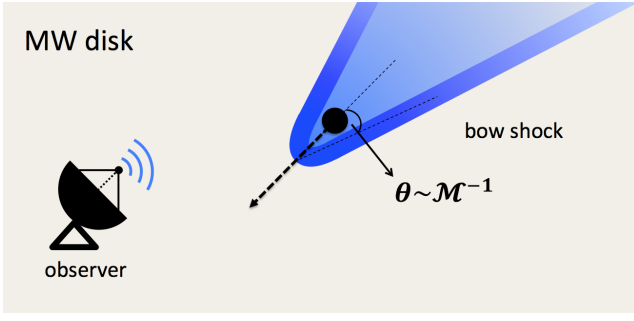}
\caption{\label{bsc}Sketch of the bow shock geometry
around a BH crossing the gaseous MW disk.}
\label{fig1}
\end{figure}

As a floating BH travels through the MW disk supersonically,
a bow shock is formed with a half opening angle 
$\theta\sim\mathcal{M}^{-1}$ \citep{shu1992,kim2009},
where the Mach number is given by
$\mathcal{M}=\vbh/\cs\approx17.0\,\v2\,\t4^{-1/2}$ (see Fig.\ref{bsc}).
The electrons accelerated in the shock produce non-thermal radiation that can be detected.
\subsection{Non-thermal spectrum}
\subsubsection{Single electron}
Next, we calculate synchrotron emission from the shock accelerated electrons around the BH.
We adopt $\n0=1$ and $\t4=1$
in the numerical examples that follow.
From the Rankine-Hugoniot jump conditions for a strong shock
the density of the shocked gas is
$n_{\rm s}\approx\left(\Gamma+1\right)\,\nism/\left(\Gamma-1\right)=4\,\nism$,
whereas its temperature is,
$T_{\rm s}=[(\Gamma+1)+2\Gamma(\mach^{2}-1)]
[(\Gamma+1)+(\Gamma-1)(\mach^{2}-1)]T/(\Gamma+1)^{2}\mach^{2}$.
The magnetic field can by obtained by assuming 
a near-equipartition of energy 
$U_{\rm B}=B^{2}/8\pi=\xi_{\rm B}\,n_{\rm s}\,kT_{\rm s}$,
where $\xi_{\rm B}$ is the fraction of thermal energy carried by the magnetic field.
Thus, the magnetic field behind the shock is given by
\begin{equation}
B\approx35\,\xi_{\rm B,-1}^{1/2}\,\n0^{1/2}\,\t4^{1/2}\;\mu\rm G\; ,
\end{equation}
where $\xi_{\rm B,-1}=\left(\xi_{\rm B}/0.1\right)$. 
We adopt $\xi_{\rm B,-1}=1$ in what follows in analogy with
supernova (SN) remnants \citep{chevalier1998}.

For a single electron with Lorentz factor $\gamma$,
the peak of its synchrotron radiation is at a frequency
$\nu_{\rm syn}=4.2\,\b-5\,\g4^{2}\; \rm GHz$,
where $\g4=\left(\gamma/10^{4}\right)$ and $\b-5=\left(B/10^{-5}\,\rm G\right)$.
The total emitted power per unit frequency is given by \citep{rybicki1979}
\begin{equation}
P(\nu)=\frac{\sqrt{2}\,e^{3}B}{\mel c^{2}}F(x)\; ,
\end{equation}
where $F(x)\equiv x\int^{\infty}_{x} K_{5/3}(\xi)\,d\xi$,
$K_{5/3}(x)$ is the modified Bessel function of $5/3$ order,
$x=\nu/c_{1}B\gamma^{2}$, $c_1=\sqrt{6}\,e/4\pi \mel c$,
$c$ is the speed of light
and $\mel$, $e$ are the electron mass and charge respectively.
The pitch angle is assumed to be uniformly distributed.

The total power from synchrotron emission of a single electron is given by
\citep{rybicki1979}
\begin{equation}
P_{\rm syn}=\frac{4}{9}r_{\rm 0}^{2}c\beta^{2}\gamma^{2}B^{2}=
2.5\times10^{-18}\b-5\,\vghz\; \rm erg\,s^{-1}\; ,
\end{equation}
where $r_{\rm 0}=e^{2}/\mel c^{2}$ is the classical radius of the electron and
$\vghz=\left(\nu_{\rm syn}\rm/GHz\right)$.
We estimate the cooling time to be $t_{\rm cool}=\gamma mc^{2}/P_{\rm syn}=
5.0\times10^{7}\b-5^{-3/2}\vghz^{-1/2} \;\rm yr$ for $\v2=1$.
Since most of the emission is near the head of the Mach cone,
we compare the cooling timescale with the dynamical timescale,
which is given by 
$t_{\rm dyn}=\rbondi/\vbh\approx53\,\m5\v2^{-3}\;\rm yr$.
For the emission frequencies of interest,
the cooling time is much longer than the lifetime of the shock.
\subsubsection{Power-law distribution of electrons}
\begin{figure}
\includegraphics[angle=0,width=\columnwidth]{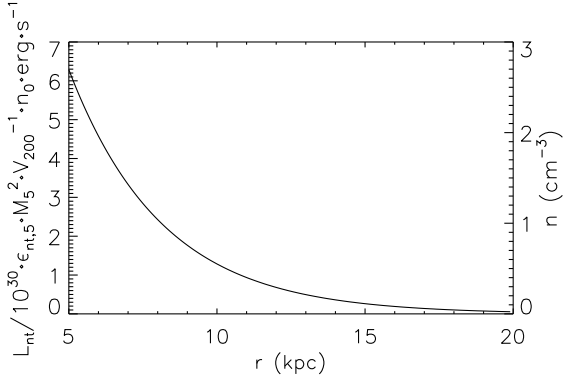}
\caption{\label{gden}
Gas density, $\nism(z=0)$, and non-thermal luminosity
in units of $10^{30}\ent5\m5^{2}\,\v2^{-1}\,\n0$ in the midplane of the MW disk.}
\label{fig2}
\end{figure}

Next we consider a broken power law distribution of electrons generated via Fermi acceleration:
\begin{equation}
N(\gamma)\,d\gamma=K_0\gamma^{-p}\left(1+\frac{\gamma}{\gb}\right)^{-1}
\;(\gmin \leq \gamma \leq \gmax),
\end{equation}
where $K_0$ is the normalization factor in electron density distribution,
$p$ is the electron power law distribution index, and 
$\gb$, $\gmin$, $\gmax$ are the break, minimum and maximum Lorentz factor respectively.
The break in the power law is due to synchrotron cooling.
The total synchrotron power can be written as,
\begin{equation}
\begin{split}
\lnt&=\epnt L_{\rm kin}=
\int^{\gmax}_{\gmin} P_{\rm syn}N(\gamma)\,d\gamma\\
&=2.3\times10^{30}\ent5\,\m5^{2}\,\v2^{-1}\,\n0\;\rm erg\,s^{-1}\; ,
\end{split}
\end{equation}
where $\ent5=(\epnt/5\%)$ is the fraction of electrons accelerated
to produce non-thermal radiation.
The normalization constant $K_0$ is obtained from the relation
$K_0=\lnt/\int^{\gmax}_{\gmin} P_{\rm syn}\gamma^{-p}\,d\gamma$.
Observations imply that the ISM density distribution in the MW disk midplane
can be roughly described by the form 
\citep{spitzer1942, kalberla2009},
\begin{equation}
\nism(r,z)=n_{\rm c}e^{-(r-\rsun)/R_{\rm n}}{\rm sech}^{2}\left[\frac{z}{\sqrt{2}z_{0}(r)}\right]\;,
\end{equation}
where $r$ and $z$ are the radial and vertical coordinates relative to the disk midplane,
$n_{\rm c}=0.9\,\rm cm^{-3}$, $R_{\rm n}=3.15\,\kpc$ and
$z_{0}(r)$ is the scale height at $r$, given by 
$z_{0}(r)=h_{0}\,e^{(r-\rsun)/r_{0}}$
with $h_{0}=0.15\,\rm kpc$, $\rsun=8.5\,\rm kpc$ and $r_{0}=9.8\,\rm kpc$ 
\citep{kalberla2009}.
The gas density and non-thermal luminosity as a function of radius 
in the MW disk midplane are shown in Figure \ref{gden}.

The electron acceleration time scale is given by
$\tacc=\xacc\gamma \mel c^{3}/eB\vbh^{2}$,
where $\xacc$ is a dimensionless constant of order unity (Blandford \& Eichler 1987).
The upper limit of the Lorentz factor $\gmax$ can be obtained 
by equaling the acceleration and cooling timescale of electrons,
$\tacc=\tc$, giving
\begin{equation}
\gmax=\frac{3\mel c\vbh}{2\,\xacc^{1/2}B^{1/2}e^{3/2}}
=2.5\times10^{7}\;\v2\,\b-5^{-1/2}\; .
\end{equation}
Since the time the gas stays in the shocked region for the electrons to be
accelerated is roughly the dynamical timescale,
an additional constraint on $\gmax$ can be obtained by 
equating the acceleration timescale of electrons 
and the dynamical time, $\tacc=\tdyn$, giving
\begin{equation}
\gmax=\frac{eB\vbh\rbondi}{\xacc\mel c^3}
=1.2\times10^5\,\b-5\m5\v2^{-1}\;.
\end{equation}
We will adopt this tighter constraint on $\gmax$ in the following calculation.
The emission frequency associated with $\gmax$ is
$\nu_{\rm max}=3\,\gmax^{2}eB/4\pi \mel c=
4.2\times10^{11}\,\b-5\gamma_{\rm max,5}^{2}\,\rm Hz$,
where $\gamma_{\rm max,5}=(\gmax/10^{5})$.
The break Lorentz factor can be obtained by equaling the cooling
and the dynamical time, giving
$\gb=5.0\times10^{9}\b-5^{-2}\,\m5^{-1}\v2^{3}$ and the corresponding frequency 
$\nu_{\rm b}=4.2\times10^{10}\,\b-5\gamma_{\rm b,9}^2\rm\,GHz$,
where $\gamma_{\rm b,9}=(\gb/10^{9})$.
The value of $\nu_{\rm b}$ is above the frequency range of interest here
and does not affect the observable synchrotron spectrum.

\begin{figure}
\begin{centering}
\includegraphics[angle=0,scale=0.5]{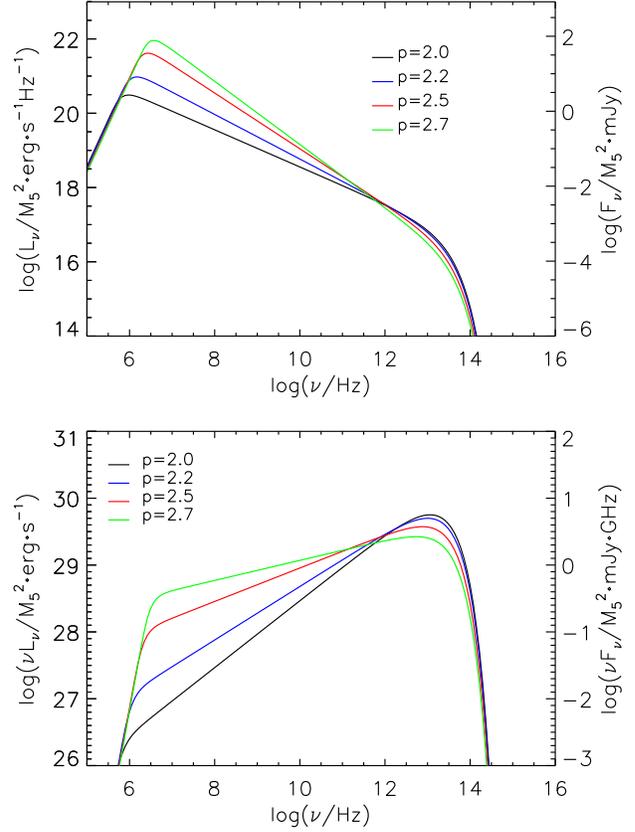}
\caption{\label{syn}Synchrotron power and flux
from non-thermal electrons accelerated by the bow shock of floating BHs,
in units of $\m5^{2}$, for
$\n0=1$, $\v2=1$, $\lnt=3.0\times10^{30}\; \rm erg\,s^{-1}$, 
$\b-5=3.5$, $\gmin\sim 1$ and $\gmax\sim4.2\times10^{5}$.
The upper panel shows synchrotron flux 
while the lower panel shows the corresponding power.
The left label of the vertical axis marks synchrotron luminosity per unit frequency
(upper panel) or power per $\log \nu$ (lower panel) 
while the right one marks the corresponding flux at a distance of $d=10\rm\;kpc$.
The black, blue, red and green lines correspond to power-law indices
$p=2.0,\;2.2,\;2.5,\;2.7$ respectively in the electron energy distribution.
Synchrotron self-absorption is significant at a frequency $\le\rm\,MHz$
and the cooling break corresponds to a frequency 
$\sim10^{19}\rm\,Hz$, which are outside the frequency range of interest.
}
\label{fig3}
\end{centering}
\end{figure}

The emissivity and absorption coefficients are given by \citep{rybicki1979}
\begin{equation}
\jv=c_{2}B\int^{\gmax}_{\gmin} F(x)
N(\gamma)\,d\gamma\; ,
\end{equation}
\begin{equation}
\av=-c_{3}B\frac{1}{\nu^{2}}\int^{\gmax}_{\gmin} \gamma^{2}
\frac{d}{d\gamma}\left[\frac{N(\gamma)}{\gamma^{2}}\right]
F(x)\,d\gamma\; ,
\end{equation}
where 
$c_2=\sqrt{2}e^{3}/4\pi \mel c^{2}$ and $c_3=\sqrt{2}e^{3}/8\pi \mel^{2}c^{2}$.
From the radiative transfer equation, the specific intensity is given by \citep{rybicki1979}
\begin{equation}
I_{\nu}=\frac{\jv}{\av}\left(1-e^{-\tv}\right)\; ,
\end{equation}
where $\tv$ is the optical depth.
The synchrotron luminosity and corresponding flux 
at a distance $d=10\;\rm kpc$ from the observer are plotted in Figure \ref{syn}.
\subsection{Emission from the vicinity of the BH}
Next we estimate the emission from the vicinity of the BH through a hot
accretion flow \citep{narayan1994}.
The Bondi accretion rate is given by 
$\mbondi=9.1\times10^{17}\m5^{2}\,\n0\v2^{-3}\;\rm g\,s^{-1}$
\citep{armitage1999}, and 
the Eddington accretion rate can be expressed as 
$\medd=L_{\rm Edd}/0.1c^{2}=1.39\times10^{23}\,\m5\;\rm g\,s^{-1}$.
We estimate the total luminosity in a radiatively inefficient accretion flow (RIAF) as,
\begin{equation}
\lbh=\eta\dot{M}c^2
=5.4\times10^{31}\zetaa^{2}\,\m5^{3}\,\n0^{2}\,\v2^{-6}\;\rm erg\,s^{-1}\;,
\end{equation}
where $\eta\approx0.1\left(\dot{M}/0.1\medd\right)$
is the efficiency of converting matter to radiation 
for $\dot{M}\le0.1\medd$ \citep{narayan2008} and
$\zeta=\dot{M}/\mbondi=10\,\zetaa$ is the accretion rate in units of $\mbondi$.
The BH accretion would produce X-ray emission which is not expected from
the bow shock spectrum in Fig.\ref{syn}.
Since $\lbh\propto\m5^{3}$,
the accretion luminosity from interstellar medium accretion onto 
stellar mass BHs is negligible compared to our souce \citep{fujita1998}.

It is possible that an outflow would be formed near the BH. 
The outflow would produce a shock at a radius $\rout$,
which can be obtained from $f\dot{M}=\mout=4\pi\rout^{2}\nism\mp\vout$, 
where $f\le1$ is the fraction of the inflowing mass channelled into the outflow.
This gives,
\begin{equation}
\begin{split}
\rout&=\left(\frac{f\dot{M}}{4\pi\nism\mp\vout}\right)^{1/2}\\
&=6.8\times10^{-4}\,f^{1/2}\zetaa^{1/2}\m5\v2^{-3/2}\vouta^{-1/2}\;\rm pc\;,
\end{split}
\end{equation}
where 
$\vouta=(\vout/10^{4}\,\kms)$ is the velocity of the outflow.
For typical parameters, we find that the outflow would be bounded with $\rout\le\rbondi$.
\subsection{Observational signatures and detectability}
Observationally, the BH emission cone would appear arc-shaped,
with an angular diameter $\theta=\rbondi/d=0.22\,\d1^{-1}\m5\v2^{-2}\, \rm arcsec$,
where $\d1=\left(d/10\,\rm kpc\right)$.
The non-thermal radiation should be detectable at radio and mm/sub-mm bands.
At a frequency $\nu\sim1\rm\,GHz$, 
the synchrotron flux at a distance of $10\rm\,kpc$
is of order $0.01-10\rm\,mJy$, depending on the choice of $p$.
This flux is detectable with the Jansky Very Large Array (JVLA),
which has a complete frequency coverage from $1-50$ GHz,
with a sensitivity of $\sim5.5\,\mu\rm Jy/beam$
in a 1-hour integration and a signal to noise ratio 
$\rm S/N=1$ at $1-2$ GHz \citep{perley2011}.
At a frequency $\nu\sim10^{10}-10^{12}\rm\,Hz$ in the mm/sub-mm band,
the synchrotron flux at a distance of $10\rm\, kpc$ is of order $10-100\,\mu\rm Jy$,
which is detectable by the Atacama Large Millimeter/sub-millimeter Array (ALMA),
covering a wavelength range of $0.3-9.6 \rm \,mm$,
with an integration time of roughly $10^{4}\rm\,s$.

Morphologically, it is possible to distinguish the bow shock emission
from other radio sources such as SN remnants or HII regions.
The bow shock emission is elongated along the direction of the BH's motion,
whereas SN remnants would appear roughly circular on the sky.
There are hundreds of cometary HII regions produced by a combination of
supersonic motion of an OB-type star through dense gas and ionization of
gas down a density gradient \citep{cyganowski2003, immer2014}.
The Mach cone's opening angle can be used to 
distinguish them from the much faster floating BHs. 
The ongoing survey of the Galactic plane with JVLA \citep{nrao2014}
has the potential to separate out these HII regions.
There are far fewer confusing HII region sources at larger radius in the disk.
Other high-velocity sources are
pulsar wind nebulae \citep{gaensler2005}, hyper-velocity stars \citep{brown2006} 
and runaway stars \citep{delvalle2012, delvalle2013}.
The first type can be distinguished by observing the pulsar as well as its X-ray emission.
The last two types produce less synchrotron radiation
\citep{delvalle2012, delvalle2013}, and thus can be distinguished as well.
Globular clusters crossing the MW disk produce another class of contaminants.
Their velocity relative to the disk is much larger than 
the velocity dispersion of their stars,
so their Bondi radius is much smaller than their size.
Thus, they should not produce significant synchrotron emission.
The floating BHs are also embedded in a star cluster, but the cluster size is more compact
and its gravity is dominated by the central BH \citep{oleary2009,oleary2012}.
\section{Summary and discussion}
\label{sec:sum}
If a floating BH happens to pass through the MW disk, then
the non-thermal emission from the accelerated electrons in the bow shock 
around the BH should produce detectable signals in the radio and mm/sub-mm bands.
The radio flux $\sim 0.01-10\;\rm mJy$ is detectable by JVLA,
while the mm/sub-mm flux $\sim 10-100\,\mu\rm Jy$ is detectable by ALMA.

The density distribution of floating BHs in the MW has been studied by
O'Leary \& Loeb (2009, 2012) and by Rashkov \& Madau (2014).
High resolution simulations show that there is a BH of mass $\sim2\times10^5\msun$
within a few kpc from the Galactic center \citep{rashkov2014}.

Observations of the Galactic disk can be used to infer $\n0$ and $\t4$.
The BH speed $\vbh$ can then be estimated from the Mach cone angle.
The maximum Lorentz factor $\gmax$ can be inferred from 
the peak of the synchrotron spectrum.
This, in turn, yields $\b-5$ based on Eq.(3).
From the slope of the synchrotron spectrum, 
the power law index $p$ can be estimated.
Finally, with the above parameters constrained,
the synchrotron flux can be used to calibrate $\mbh$.
The above interpretation can be verified by 
observing the properties of the star cluster carried by the floating BHs
\citep{oleary2009, oleary2012}.
The diffuse X-ray emission from the BH and
synchrotron emission from the bow shock is supplemented by
stellar emission from the star cluster around it.
Since the total mass of the star cluster is much smaller than $\mbh$,
gravity is dominated by the BH, and thus the stars do not effect the bow shock.
One can measure $\mbh$ spectroscopically from the velocity dispersion of the stars
as a function of distance from the BH, 
and verify consistency with the synchrotron flux estimate.
\section*{Acknowledgements}
We thank Jonathan Grindlay, James Guillochon, Ramesh Narayan, Mark Reid and Lorenzo Sironi
for helpful comments on the manuscript.
We thank Piero Madau for providing the data from \textit{Via Lactea II} simulation.
This work was supported in part by NSF grant AST-1312034.
%
%

\label{lastpage}
\end{document}